\documentclass[prb,twocolumn,showpacs,preprintnumbers,amsmath,amssymb,superscriptaddress,bibnotes,floatfix]{revtex4}

\usepackage{color}
\usepackage{graphicx}
\usepackage{dcolumn}
\usepackage{bm}
\usepackage{mathrsfs}

\begin{document}

\title{Enhanced anisotropic spin fluctuations below tetragonal-to-orthorhombic transition\\ in LaFeAs(O$_{1-x}$F$_x$) probed by $^{75}$As and $^{139}$La NMR}

\author{Yusuke Nakai}\thanks{Present address: Department of Physics, Graduate School of Science and Engineering, Tokyo Metropolitan University, Tokyo 192-0397, Japan}
\email{nakai@tmu.ac.jp}
\author{Shunsaku Kitagawa}
\author{Tetsuya Iye}
\author{Kenji Ishida}
\affiliation{Department of Physics, Graduate School of Science, Kyoto University, Kyoto 606-8502, Japan,}
\affiliation{TRIP, JST, Sanban-cho Building, 5, Sanban-cho, Chiyoda, Tokyo 102-0075, Japan,}

\author{Yoichi Kamihara}\thanks{Present address: Department of Applied Physics \& Physico-Informatics, Faculty of Science \& Technology, KEIO University, Yokohama 223-8522, Japan,} 
\affiliation{TRIP, JST, Sanban-cho Building, 5, Sanban-cho, Chiyoda, Tokyo 102-0075, Japan,}
\author{Masahiro Hirano}
\affiliation{ERATO-SORST, JST, Frontier Collaborative Research Center, Tokyo Institute of Technology, Yokohama 226-8503, Japan,} 
\affiliation{Frontier Research Center, Tokyo Institute of Technology, Yokohama 226-8503, Japan,}
\author{Hideo Hosono}
\affiliation{ERATO-SORST, JST, Frontier Collaborative Research Center, Tokyo Institute of Technology, Yokohama 226-8503, Japan,} 
\affiliation{Frontier Research Center, Tokyo Institute of Technology, Yokohama 226-8503, Japan,}
\affiliation{Materials and Structures Laboratory, Tokyo Institute of Technology, Yokohama 226-8503, Japan}

\date{\today}

\begin{abstract}
$^{75}$As and $^{139}$La NMR results of LaFeAs(O$_{1-x}$F$_x$) ($x$=0, 0.025, and 0.04) were reported. 
Upon F-doping, the tetragonal-to-orthorhombic structural phase transition temperature $T_S$, antiferromagnetic transition temperature $T_N$ and internal magnetic field $\mu_0H_{\rm int}$ are gradually reduced for $x<0.04$. 
However, at $x=0.04$, $T_N$ is abruptly suppressed to be 30 K along with a tiny $\mu_0H_{\rm int}$, which is distinct from the continuous disappearance of the ordered phases in the Ba122 systems of Ba(Fe,Co)$_2$As$_2$ and BaFe$_2$(As,P)$_2$. 
The anisotropy of the spin-lattice relaxation rate $T_1^{-1}$, $(T_1)^{-1}_{H\parallel ab}/(T_1)^{-1}_{H\parallel c}$, in the paramagnetic phase of $x = 0$ and 0.025 is constant ($\sim 1.5$), but increases abruptly below $T_S$ due to the enhancement of $(T_1)^{-1}_{H\parallel ab}$ by the slowing down of magnetic fluctuations. 
This indicates that the tetragonal-to-orthorhombic structural distortion enhances the anisotropy in the spin space via magnetoelastic coupling and/or spin-orbit interaction.
\end{abstract}

\maketitle

In the ``1111" iron-arsenide $R$FeAsO ($R$ = rare earth elements), a tetragonal-to-orthorhombic structural phase transition occurs at $T_S$, 
while a stripe-like antiferromagnetic (AF) ordering occurs at a slightly lower temperature at $T_N$.~\cite{PaglioneReview} 
In contrast, in the low-$T_c$ iron-phosphide superconductor LaFePO,~\cite{KamiharaFeP} there are neither AF ordering nor structural phase transition, 
suggesting that they may be essential for the high-$T_c$ superconductivity in iron-based superconductors.
Understanding the connection between magnetism and structural distortion in LaFeAsO is the first important step toward clarifying their relation to the high-temperature superconductivity. 
Until now, however, the relationship between the structural and magnetic phase transition is an open question.
The structural transition has been suggested to arise from various kinds of electron order such as spin nematic ordering~\cite{CFangPRB2008,CXuPRB2008} and orbital order.~\cite{LeePRLOrbitalorder2009,KrugerPRBOrbitalOrder2009} 
Recent experiments on detwined samples found strong anisotropy in the electronic properties of the orthorhombic phase, suggesting an underlying electronic nematic state.~\cite{ChuangScience2010,ChuScience2010}
It has been proposed that the structural distortion involves a change in the orbital configuration.~\cite{ShimojimaPRL}

In this paper, we report $^{75}$As and $^{139}$La NMR results in LaFeAs(O$_{1-x}$F$_x$) ($x=$0, 0.025, and 0.04). F-doping suppresses $T_N$ and $T_S$ rapidly at $x\sim0.04$, and the first-order like transition against F content is realized at $x\sim0.04$, which is consistent with previous reports. Such a sudden disappearance of the ordered phases is in acute contrast with the Ba122-type iron-based superconductors where the ordered phase is continuously suppressed by chemical substitution. 
In the paramagnetic phase, $(T_1T)^{-1}$ exhibits anisotropic behavior [$(T_1T)^{-1}_{H\parallel ab}$, measured in the magnetic field perpendicular to the $c$ axis, is 1.5 times larger than $(T_1T)^{-1}_{H\parallel c}$, measured in the magnetic field parallel to the $c$ axis,] and becomes more anisotropic below $T_S$, indicative of the presence of magnetoelastic coupling and/or spin-orbit interaction. 

Polycrystalline samples of LaFeAs(O$_{1-x}$F$_x$) ($x=$0, 0.025, and 0.04) were grown by solid state reactions.~\cite{KamiharaFeAs} 
Powder x-ray diffraction indicates that they are mostly of single-phase and contain only a small amount of FeAs ($\sim$1\%). 
The $x$ values were determined by lattice constants using Vegard's volume law. The error in the doping level is evaluated to be less than 0.5 \%. 
Previous electrical resistivity, susceptibility,~\cite{KamiharaFeAs} specific heat,~\cite{KohamaFeAsPRB2008} synchrotron x-ray diffraction,~\cite{Nomura} M\"ossbauer~\cite{KitaoJPSJ2008} and NMR measurements~\cite{NakaiJPSJ2008,NakaiNJP2009} showed $T_S\simeq165$ K and $T_N\simeq142$ K for LaFeAsO. 
A superconducting (SC) transition determined by zero resistivity was observed at $T_c = 16.3$ K in $x = 0.04$, but was not observed in $x = 0.025$. 

First, we show the evolution of $T_N$ and the internal magnetic field $\mu_0H_{\rm int}$ upon F doping revealed by $^{139}$La NMR spectra. 
Figures~1~(a) and (b) display $^{139}$La NMR spectra of our powdered samples of $x = 0.025$ and 0.04, which are obtained by sweeping magnetic field at a fixed frequency of 40.5 MHz. 
$^{139}$La NMR spectra of $x = 0.025$ becomes broadened due to the appearance of $\mu_0H_{\rm int}$ below approximately 120 K, and three peaks were observed at 10 K (Fig.~1~(a)). 
These peaks were assigned as follows; one peak at $\mu_0H\sim 6.8$ T corresponds to $H\parallel ab$ and the remaining two split peaks correspond to $H\parallel c$ since the ordered Fe moments induce $H_{\rm int}$ parallel to the $c$ axis at the La site.~\cite{KitagawaBaFe2As2} 
The spectrum at 10 K is reproduced consistently by powder lineshape simulations based on the above assignment. 
From the simulations, we found that the two peaks around 6.6 and 7 T correspond to the first satellite peaks of $^{139}$La NMR spectrum split due to the appearance of the internal magnetic field. 
The internal magnetic field $\mu_0H_{\rm int}$ determined by the splitting of the two peaks shown by the arrows in Fig.~1~(a) is plotted in Fig.~1~(c). 
In the undoped LaFeAsO, we previously observed $\mu_0H_{\rm int}$ determined from $^{139}$La NMR spectra below $T_N\simeq142$ K,~\cite{NakaiJPSJ2008} which is shown by the closed circles in Fig.~1~(c). The dashed line in Fig.~1~(c) is a fitting to the expression, $M(T)/M_0 = [1-(T/T_N)]^{0.15}$, where $M$ is the ordered moment. 
The growth of $M(T)/M_0$ below $T_N$ is much steeper than 3D mean field value (0.5), and the transition to the antiferromagnetic state is almost first-order like. 
This is consistent with the previous $\mu$SR experiments as shown by the open circles.~\cite{LuetkensmuSR} 
In contrast, for $x=0.04$, there is no obvious splitting but only a slight broadening of the NMR spectrum appears as shown in Fig.~1~(b). 
Thus, the half width at half maximum (HWHM) of the central peak of $x = 0.04$ is plotted in Fig.~1~(c). 
The gradual increase in HWHM suggests the occurrence of magnetic order with reduced moment below 30 K for $x = 0.04$. 
To confirm this magnetic ordering, we measured NMR spectrum at the As site for $x = 0.04$, since the hyperfine coupling at the As site is larger than that at the La site, and thus larger broadening is expected. 
Indeed, a broadening of HWHM of the central peak was observed in the $^{75}$As NMR spectrum as shown in Fig.~2. 
In addition, a peak in $^{75}$As $(T_1T)^{-1}$ at ~30 K due to the slowing down of AF spin fluctuations as shown below (see Fig.~5~(a)) also evidences a magnetic ordering at $x = 0.04$. 
From the broadening behavior of the central peak of the $^{139}$La NMR spectra, we estimated the upper bound for the internal magnetic field for $x = 0.04$ as $\mu_0H_{\rm int}\le0.01$ T, which is much smaller than $\mu_0H_{\rm int}$ of $\simeq0.38$ T for $x = 0.025$. 
From the F doping dependence of $\mu_0H_{\rm int}$ at low temperatures, we found that the ordered moments abruptly decrease at around $x = 0.04$, which is consistent with previous results.~\cite{LuetkensNatMatLaFeAsOF,HuangPRBLaFeAsOF,QureshiPRB2010}

\begin{figure}[tb]
\begin{center}
\includegraphics[width=8cm]{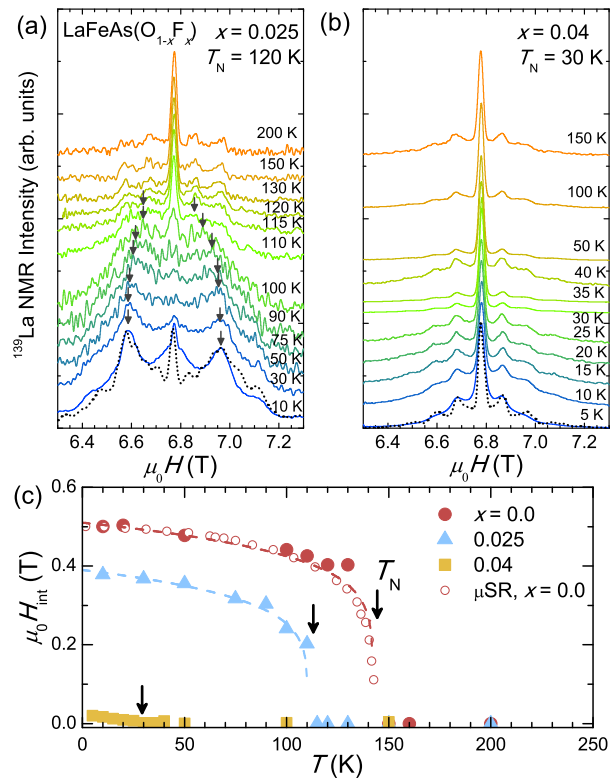}
\end{center}
\caption{(Color online) $^{139}$La NMR spectra for (a) $x=$0.025 and (b) 0.04. The arrows indicate the splitting due to the internal magnetic field. The pair of the peaks at 6.6 T and 7 T corresponds to $H\parallel c$, whereas the peak near 6.8 T corresponds to $H\parallel ab$ due to partial alignment of the $x=0.025$ powder sample. 
For $x=0.04$, there is no obvious splitting but only a slight broadening is observed, indicating a tiny ordered moment at $x=0.04$. 
The dotted lines are simulated spectra for (a) a partially aligned powder pattern with $\mu_0H_{\rm int} = 0.38$ T and $\nu_Q = 1.3$ MHz, and (b) a powder pattern with $\mu_0H_{\rm int} = 0.01$ T and $\nu_Q = 1.2$ MHz.
(c) $T$-dependence of the internal magnetic field at the La site determined by the $^{139}$La NMR spectra. 
}
\end{figure}

\begin{figure}[tb]
\begin{center}
\includegraphics[width=8cm]{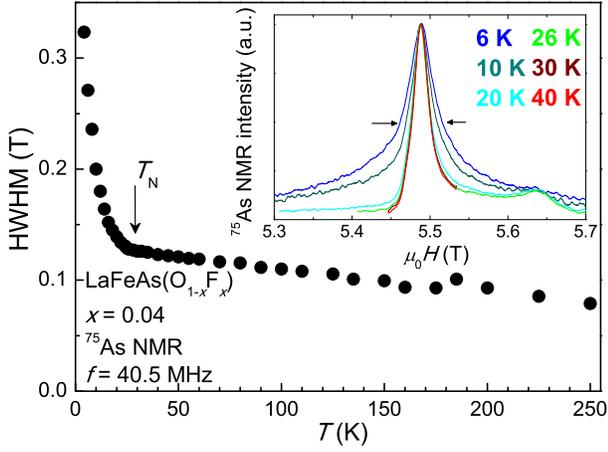}
\end{center}
\caption{(Color online) Half width at half maximum of the center line of $^{75}$As NMR spectra for $x = 0.04$, which are shown in the inset.} 
\end{figure}

\begin{figure}[tb]
\begin{center}
\includegraphics[width=7.5cm]{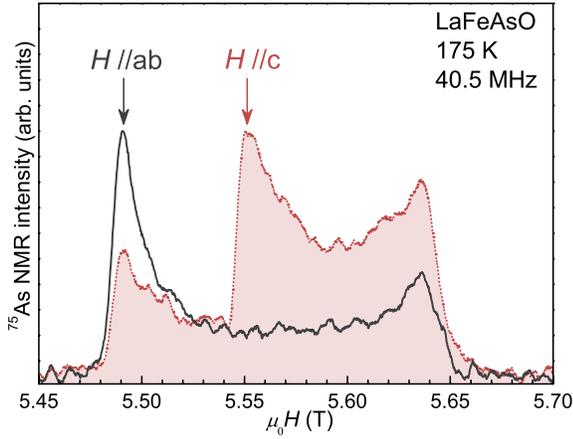}
\end{center}
\caption{(Color online) $^{75}$As NMR spectra of the central transition for an aligned powder sample of LaFeAsO obtained by sweeping magnetic field perpendicular to the plane where the $c$ axis is randomly distributed (solid line) and parallel to the plane (dotted line) above $T_S$.} 
\end{figure}
\begin{figure}[tb]
\begin{center}
\includegraphics[width=8cm]{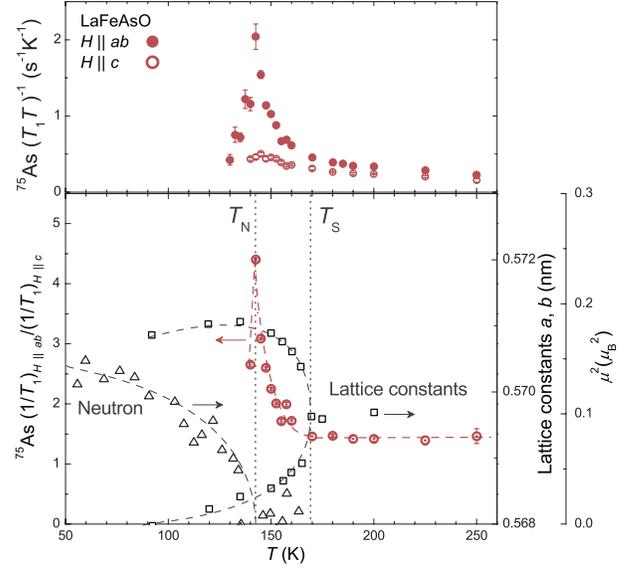}
\end{center}
\caption{(Color online) $T$-dependence of $^{75}$As $(T_1T)^{-1}$ for $H\parallel ab$ and $c$ (upper panel), and the anisotropy of $T_1$ defined as $(T_1)^{-1}_{H\parallel ab}/(T_1)^{-1}_{H\parallel c}$ along with the lattice constants and the square of the ordered moment~\cite{Cruz} (lower panel). 
The anisotropy starts to increase below $T_S$ as inferred from the change in the lattice constants.~\cite{Nomura}} 
\end{figure}
Next, we discuss anisotropic behavior of $T_1$ in LaFeAs(O$_{1-x}$F$_x$) revealed by $^{75}$As NMR measurements. 
In order to obtain the anisotropy of $T_1$, we aligned our powdered samples. All our samples were mixed with Stycast 1266 epoxy, and allowed to cure in 5.5 T at room temperature. Due to the anisotropy of magnetic susceptibility, the powdered samples were partially aligned with the $c$ axis perpendicular to the applied field (``two-dimensional alignment"). 
Figure~3 displays $^{75}$As NMR spectra of the aligned powder of LaFeAsO above $T_S$ obtained by sweeping magnetic field perpendicular to the plane where the $c$ axis is randomly distributed (solid line) and parallel to the plane (dotted line). A peak at about 5.55 T corresponding to $H\parallel c$ was clearly observed, and similar spectra were observed for $x=0.025$ and 0.04. $^{75}$As $1/T_1$ was measured at each central line for $H\parallel ab$ and $c$. 
$T_1$ was measured with a saturation recovery method, and was fitted with a single component in a whole measured temperature range except in $x$ = 0.04 below 30 K.~\cite{NakaiJPSJ2008}

As shown in the upper panel of Fig.~4, $(T_1T)^{-1}$ for $H\parallel ab$ increases rapidly on cooling and exhibits a pronounced peak at $T_N$, which is consistent with our previous $^{139}$La-NMR results.~\cite{NakaiJPSJ2008} 
In contrast, $(T_1T)^{-1}$ for $H\parallel c$ increases gradually on cooling and exhibits a tiny bump at $T_N$. 
The anisotropic behavior of $(T_1T)^{-1}$ is clearly seen in the lower panel of Fig.~4 where the anisotropy of $T_1$ defined as $(T_1)^{-1}_{H\parallel ab}/(T_1)^{-1}_{H\parallel c}$ is plotted. 
The anisotropy is constant at high temperatures, but starts to increase at around 170 K, indicating that fluctuating hyperfine fields at the $^{75}$As site become anisotropic below  approximately 170 K. Interestingly, the tetragonal-to-orthorhombic structural phase transition temperature $T_S$ inferred from the lattice constants coincides with the temperature at which the anisotropy of $T_1$ starts to develop. This indicates that the structural phase transition has a strong impact on the spin dynamics, and that measurements of the anisotropy can be utilized as an indicator of the structural phase transition.

Similar anisotropic behavior of $T_1$ was observed for $x = 0.025$ as shown in Fig.~5.
A pronounced peak was observed for $H\parallel ab$ at $\sim$120 K for $x=0.025$, and the anisotropy of $T_1$ starts to increase below approximately 140 K which we identified as $T_S$ for $x=0.025$. 
In fact, previous specific heat measurements have detected an anomaly around 140 K.~\cite{KohamaFeAsPRB2008}
By contrast, although a peak related to $T_N$ was observed in $(T_1T)^{-1}$ for $H\parallel ab$ at $\sim$ 30 K in $x=0.04$, such a significant increase in the anisotropy of $T_1$ was not observed for $x=0.04$, but the anisotropy increases very gradually on cooling, preventing us from determining $T_S$. 
The absence of a pronounced signature of $T_S$ may be naturally understood by the distribution of $T_S$ because the $x=0.04$ sample locates at around the phase boundary.~\cite{LangPRL2010} 

Here, we analyze the $T_1$ data following previous studies.~\cite{KitagawaBaFe2As2,SKitagawaAnisotropy} 
Hyperfine fields at the $^{75}$As site, $\bm{H}_{\rm hf}^{\rm As}$, can be described as the sum of contributions from the four nearest neighbor Fe spins, 
\begin{equation}
\bm{H}_{\rm{hf}}^{\rm{As}} = \sum_{i=1}^{4} \bm{A_{i}} \cdot \bm{S}_{i}
            = \Tilde{A} \bm{S},
\end{equation} 
where $\bm{S}_{i}$ is the Fe electron spin at the $i$-th Fe site, $\bm{A_{i}}$ is the hyperfine coupling tensor to the electron spin at the $i$-th Fe site, and $\Tilde{A}$ is the hyperfine coupling tensor ascribed to the four nearest neighbor Fe electron spins. 
As shown previously,~\cite{KitagawaBaFe2As2,SKitagawaAnisotropy} $\tilde{A}$ can be described as follows in the orthorhombic notation: 
\begin{eqnarray}
\tilde{A}=\left(
\begin{array}{ccc}
A_a&C &B_1 \\
C&A_b &B_2 \\
B_1&B_2 &A_c \\
\end{array}
\right).
\end{eqnarray} 
$A_i$ originates from paramagnetic (PM) correlations along $i$-axis. $B_{1[2]}$ originates from the stripe ($\pi$, 0) [(0, $\pi$)] AF correlations and $C$ originates from the checkerboard ($\pi$, $\pi$) AF correlations. 
As discussed in the previous papers, if the stripe correlations dominates the system like LaFeAsO,~\cite{IshikadoLaFeAsO} 
the ratio of $(T_1)^{-1}_{H\parallel ab}/(T_1)^{-1}_{H\parallel c}$ can be written as,
\begin{equation}
\frac{(T_1)^{-1}_{H\parallel ab}}{(T_1)^{-1}_{H\parallel c}} = 
\left|\frac{S_a(\omega_{\rm res})}{S_c(\omega_{\rm res})}\right|^2 + \frac{1}{2}
\label{T1anisotropy}
\end{equation}
where $(T_1)^{-1}_{H\parallel ab} = \frac{(T_1)^{-1}_{H\parallel a} + (T_1)^{-1}_{H\parallel b}}{2}$ and $|S(\omega)|^{2}$ denotes the power spectral density of a time-dependent random variable $S(t)$. 
The ratio becomes 1.5 if the Fe spin fluctuations are isotropic ($|S_a|=|S_c|$) while the ratio becomes much larger than 1.5 if Fe spin fluctuations are highly anisotropic ($|S_a|\gg|S_c|$). 
Note that the ratio of 1.5 suggesting the presence of the stripe AF correlation has been reported in various iron-based superconductors, which seems to be a common feature of the superconductors.~\cite{SKitagawaAnisotropy,SKitagawaZnPRB2011,Matano(BaK)122EPL2009,ZhangBa0.3K0.7Fe2As2PRB2010,ZLi-LiFeAsJPSJ2010,LMa-LiFeAsPRB2010}

The anisotropy of $T_1$ is approximately 1.5 at high temperatures for all the samples, and increases suddenly below $T_S$ as shown in Fig.~5~(b). 
The above analysis indicates that the stripe AF fluctuations with the isotropic Fe spin component in the spin space are present above $T_S$, and that the spin component becomes anisotropic below $T_S$; Fluctuations of Fe spins in the $ab$-plane become more significant than along $c$-axis below $T_S$. 
This indicates that the tetragonal-to-orthorhombic structural transition enhances the anisotropy of the spin space via strong magnetoelastic coupling and/or spin-orbit interaction. 
\begin{figure}[tb]
\begin{center}
\includegraphics[width=7.5cm]{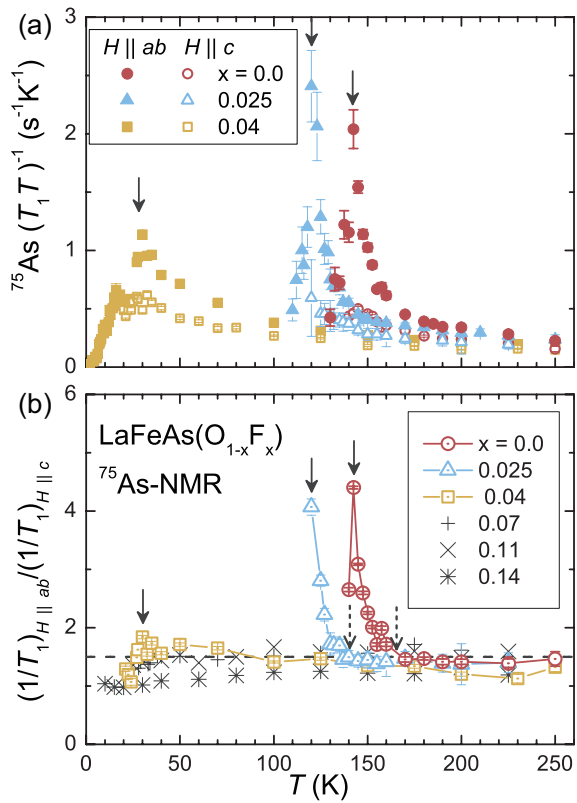}
\end{center}
\caption{(Color online) $T$-dependence of (a) $^{75}$As $(T_1T)^{-1}$ for $H\parallel ab$ and $H\parallel c$, and of (b) the anisotropy of $T_1$ defined as $(T_1)^{-1}_{H\parallel ab}/(T_1)^{-1}_{H\parallel c}$ for LaFeAs(O$_{1-x}$F$_x$) ($x=$0, 0.025, and 0.04) along with our previous NMR results for $x = 0.07$, 0.11 and 0.14.~\cite{SKitagawaAnisotropy} The solid (dotted) arrows indicate $T_N$ ($T_S$).}
\end{figure}
A similar abrupt increase in the anisotropy of $T_1$ below a tetragonal-to-orthorhombic structural transition temperature was observed in NaFeAs with the ``111" structure,~\cite{KitagawaNaFeAs,LMa-NaFeAsPRB2011} which undergoes the structural transition followed by an AF transition at a lower temperature as in LaFeAsO. 
Thus, the sudden increase in the anisotropy of $T_1$ below $T_S$, indicating the enhanced anisotropy in the spin space, can be observed in compounds with the separation of $T_N$ and $T_S$. 
Indeed, the ``122" parent compounds, where these two transitions take place simultaneously, do not exhibit such an abrupt increase in the anisotropy of $T_1$.~\cite{KitagawaSrFe2As2} 
We note that an enhancement of the anisotropy in the spin space was not observed but the anisotropy ratio of 1.5 was observed in the superconducting LaFeAs(O$_{1-x}$F$_x$) ($x\ge0.07$) with the tetragonal structure as shown in Fig.~5~(b).~\cite{SKitagawaAnisotropy} This suggests that the stripe AF correlation is essential for superconductivity in iron-pnictide superconductors.\cite{SKitagawaAnisotropy}

In conclusion, we report $^{75}$As and $^{139}$La NMR results of the undoped and underdoped LaFeAs(O$_{1-x}$F$_x$) ($x=$0, 0.025, and 0.04). 
F-doping suppresses the AF and structural phase transition temperature very rapidly, and the first-order-like transition against F content is observed. Such a sudden disappearance of the ordered phases is in quite contrast with the Ba122-type iron-based superconductors such as Ba(Fe, Co)$_2$As$_2$ and BaFe$_2$(As, P)$_2$.~\cite{NingPRL2010,NakaiPRL2010}  
In the paramagnetic phase of $x=0$ and 0.025, $(T_1T)^{-1}$ becomes more anisotropic below the structural phase transition temperature. This indicates the presence of the strong magnetoelastic coupling and/or spin-orbit interaction in LaFeAs(O$_{1-x}$F$_x$). 

We are grateful to S. Yonezawa, Y. Maeno and H. Ikeda for fruitful discussion. 
This work was supported by the Grants-in-Aid for Scientific Research on Innovative Areas ``Heavy Electrons" (No. 20102006) from MEXT, for the GCOE Program ``The Next Generation of Physics, Spun from Universality and Emergence" from MEXT, and for Scientific Research from JSPS. 
One of the authors(HH) acknowledges the support by the FIRST program, JSPS, JAPAN.

\end{document}